\documentclass[aps,pra,twocolumn,showpacs,floatfix]{revtex4}
\usepackage{bbm,graphicx,amsfonts,amsbsy,mathrsfs}

\def\alphaZero{\alpha^{(0)}_{f,f^\prime}}
\def\alphaOne{\alpha^{(1)}_{f,f^\prime}}
\def\alphaTwo{\alpha^{(2)}_{f,f^\prime}}

\def\aX{\hat{a}^{\phantom{\dagger}}_{\mathrm{x}}}
\def\aXDagger{\hat{a}^{\dagger}_{\mathrm{x}}}
\def\aY{\hat{a}^{\phantom{\dagger}}_{\mathrm{y}}}
\def\aYDagger{\hat{a}^{\dagger}_{\mathrm{y}}}
\def\aXp{\hat{a}^{\phantom{\dagger}}_{\mathrm{x}'}}
\def\aXpDagger{\hat{a}^{\dagger}_{\mathrm{x}'}}
\def\aYp{\hat{a}^{\phantom{\dagger}}_{\mathrm{y}'}}
\def\aYpDagger{\hat{a}^{\dagger}_{\mathrm{y}'}}
\def\aMinus{\hat{a}^{\phantom{\dagger}}_{-}}
\def\aMinus{\hat{a}^{\phantom{\dagger}}_{-}}
\def\aMinusDagger{\hat{a}^\dagger_{-}}
\def\aPlus{\hat{a}^{\phantom{\dagger}}_{+}}
\def\aPlusDagger{\hat{a}^\dagger_{+}}

\def\UtTilde{\tilde{U}^{\phantom{\dagger}}_{\delta t}}

\def\F{\mathbf{F}}
\def\Fx{F_\mathrm{x}}

\def\Fz{F_\mathrm{z}}

\def\fMinus{\hat{f}^{\phantom{\dagger}}_-}
\def\fMinusSquared{\hat{f}^{2\phantom{\dagger}}_-}
\def\fPlus{\hat{f}^{\phantom{\dagger}}_+}
\def\fPM{\hat{f}^{\phantom{\dagger}}_\pm}
\def\fPlusSquared{\hat{f}^{2\phantom{\dagger}}_+}
\def\fX{\hat{f}^{\phantom{\dagger}}_\mathrm{x}}
\def\fXSquared{\hat{f}^{2\phantom{\dagger}}_\mathrm{x}}
\def\fY{\hat{f}^{\phantom{\dagger}}_\mathrm{y}}
\def\fYSquared{\hat{f}^{2\phantom{\dagger}}_\mathrm{y}}
\def\fZ{\hat{f}^{\phantom{\dagger}}_\mathrm{z}}
\def\fZSquared{\hat{f}^{2\phantom{\dagger}}_\mathrm{z}}

\def\sZero{\hat{S}^{\phantom{\dagger}}_{0}}
\def\sX{\hat{S}^{\phantom{\dagger}}_{\mathrm{x}}}
\def\sY{\hat{S}^{\phantom{\dagger}}_{\mathrm{y}}}
\def\sZ{\hat{S}^{\phantom{\dagger}}_{\mathrm{z}}}

\def\fIdent{\hat{\mathbbm{1}}^{\phantom{\dagger}}_{f}}

\def\Pf{\hat{P}_f}
\def\Pfp{\hat{P}_{f'}}

\def\beDef{\vec{\mathbf{e}}}
\def\e{\vec{\mathbf{e}}}
\def\eStar{\beDef^{*}}

\def\eMinus{\beDef^{\phantom{*}}_{-}}
\def\eMinusStar{\beDef^{*}_{-}}
\def\ePlus{\beDef^{\phantom{*}}_{+}}
\def\ePlusStar{\beDef^{*}_{+}}
\def\eZero{\beDef^{\phantom{*}}_{0}}
\def\eZeroStar{\beDef^{*}_{0}}
\def\eX{\beDef^{\phantom{*}}_{x}}

\def\eY{\beDef^{\phantom{*}}_{y}}

\def\eZ{\beDef^{\phantom{*}}_{z}}

\def\ENegative{\hat\mathbf{E}^{(-)}}
\def\EPositive{\hat\mathbf{E}^{(+)}}

\def\dC{\hat{d}}
\def\dCDagger{\hat{d}^\dagger}
\def\dVec{\hat{\mathbf{d}}}
\def\dVecDagger{\hat{\mathbf{d}}^\dagger}
\def\dMinus{\hat{d}_-^{\phantom{\dagger}}}
\def\dPlus{\hat{d}_+^{\phantom{\dagger}}}

\def\dZero{\hat{d}_0^{\phantom{\dagger}}}
\def\dMinusDagger{\hat{d}_-^\dagger}
\def\dPlusDagger{\hat{d}_+^\dagger}

\def\dZeroDagger{\hat{d}_0^\dagger}

\def\alphaTensor{\hat{\boldsymbol{\alpha}}}
\def\alphaTensorJ{\alphaTensor^{(j)}}
\def\alphaTensorZero{\alphaTensor^{(0)}}
\def\alphaTensorOne{\alphaTensor^{(1)}}
\def\alphaTensorTwo{\alphaTensor^{(2)}}

\def\hInt{\hat{H}}
\def\hIntZero{\hInt^{(0)}}
\def\hIntOne{\hInt^{(1)}}
\def\hIntTwo{\hInt^{(2)}}

\def\SNR{\mathrm{SNR}}
\def\OD{\mathrm{OD}}

\def\alphaConst#1#2{\alpha_{#1}^{#2}}
\def\kron#1#2{\delta_{#1}^{#2}}

\begin{document}

\title{Tensor polarizability and dispersive quantum measurement of multilevel atoms}
\author{JM Geremia}
\author{John K. Stockton}
\email{jks@caltech.edu}
\author{Hideo Mabuchi}
\affiliation{Physical Measurement and Control 266-33, California
Institute of Technology, Pasadena, CA 91125}

\begin{abstract}
Optimally extracting information from measurements performed on a physical system requires an accurate model of the measurement interaction.  Continuously probing the collective spin of an Alkali atom cloud via its interaction with an off-resonant optical probe is an important example of such a measurement where realistic modeling at the quantum level is possible using standard techniques from atomic physics.  Typically, however, tutorial descriptions of this technique have neglected the multilevel
structure of realistic atoms for the sake of simplification. In this paper we account for the full multilevel structure of Alkali atoms and derive the irreducible form of the polarizability Hamiltonian describing a typical dispersive quantum measurement. For a specific set of parameters, we then show that semiclassical predictions of the theory are consistent with our experimental observations of polarization scattering by a polarized cloud of laser-cooled Cesium atoms. We also derive the signal-to-noise ratio under a single measurement trial and use this to predict the rate of spin-squeezing with multilevel Alkali atoms for arbitrary detuning of the probe beam.
\end{abstract}

\date{\today}
\maketitle

%\tableofcontents

\section{Introduction}

\noindent Information gained by performing measurements on a quantum
system can reduce uncertainty about one or more of its physical
observables. It is, however, a basic property of quantum mechanics
that measurements are invasive in the sense that they necessarily
degrade one's ability to make subsequent predictions about the
values of complementary observables \cite{Braginski1992}. This type
of disturbance is often called measurement backaction, and it is a
natural consequence of the Hamiltonian coupling between a probe
(such as an electromagnetic field mode) and the system of interest.
In a special class of ``backaction evading'' experimental scenarios,
it is possible to channel the disturbance into observables that are
not dynamically coupled to the main quantities of interest. When
such measurements are performed with minimal technical imperfection
on systems whose initial preparations are sufficiently pure, which
qualifies them as what is referred to in the quantum optics
literature as quantum non-demolition (QND) measurement
\cite{Haroche1990,Haroche1999}, it is possible to create
conditionally squeezed states of the measured observable.

While measurement-induced squeezing can easily be understood in an
abstract sense, predicting the precise degree of squeezing that can
be achieved in a realistic experiment requires detailed physical
modeling of the system-probe interaction (in addition to any
operative decoherence mechanisms). Squeezed states of atomic spins
have recently emerged
\cite{Kitagawa1993,Kuzmich2000,Polzik2001,Geremia2004a} as a
versatile and robust resource for quantum information science
\cite{Sorensen2001,Sorensen2001b} and quantum metrology
\cite{Wineland1994,Geremia2003,Geremia2005b,Andre2004,Molmer2005}.
In these contexts, the degree of spin squeezing is directly linked
to entanglement measures, to achievable reductions in averaging
times for precision measurement, and to achievable improvements over
communication protocols that utilize only classical information
resources.

Theoretical analyses of measurement-induced spin squeezing typically
consider a system of $N\gg 1$ atoms whose collective spin is
described by an observable
\begin{equation}
    \hat{\mathbf{F}} = \sum_{i=1}^N {}^{(i)}\hat{\mathbf{f}},
\end{equation}
where ${}^{(i)}\hat\mathbf{f} = \cdots \otimes \hat\mathbbm{1}_{i-1}
\otimes \hat\mathbf{f} \otimes \hat\mathbbm{1}_{i+1} \otimes \cdots$
is the angular momentum (vector) operator for the $i^{th}$ atom.
Cartesian components $\hat{F_x}$, $\hat{F_y}$ and $\hat{F_z}$ follow
from this in an obvious way. Under physical conditions that preserve
permutation symmetry of the label $i$, the collective spin of an
initially polarized atomic sample can be restricted
\cite{Stockton2003} to its maximum angular momentum shell. The
associated Hilbert sub-space is spanned by eigenstates $|F,M\rangle$
of the collective spin observable $\hat{ \mathbf{F}}$ that satisfy
$\hat\mathbf{F}^2|F,M\rangle = \hbar^2 F(F+1) |F,M\rangle$, where
$F=Nf$ for atoms with individual spin $f$.

\begin{figure}[b]
\begin{center}
\includegraphics{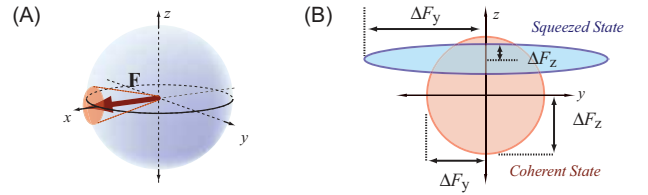}
\end{center}
\vspace{-4mm} \caption{(A) Graphical representation of the
spin-polarized atomic sample as a classical magnetization vector
with transverse quantum uncertainty. (B) Schematic of the transverse
quantum uncertainties for coherent and squeezed spin
states.\label{Figure::Polarization}} \vspace{-3mm}
\end{figure}

It is natural to conceptualize the quantum state of such a system as
a Bloch-like magnetization vector $\F\equiv[\langle\hat{F_x}\rangle,
\langle \hat{F_y}\rangle,\langle\hat{F_z}\rangle]$ plus a transverse
uncertainty $\Delta \mathbf{F}_\perp$ associated with the variances
of $\hat{F_x},$ $\hat{F_y}$ and $\hat{F_z}$ (see
Fig.~\ref{Figure::Polarization}). The transverse uncertainty
$\Delta\mathbf{F}_\perp$ can never vanish since $\hat{F_x},$
$\hat{F_y}$ and $\hat{F_z}$ do not commute; this constraint can be
interpreted to mean that we can never have perfect knowledge of the
orientation of the collective magnetization.

\begin{figure}[t]
\begin{center}
\includegraphics{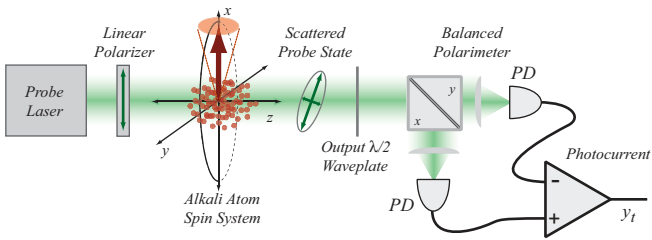}
\end{center}
\vspace{-2mm} \caption{Schematic of an experimental apparatus for
continuous measurement of collective spin in an Alkali atom sample
based on polarimetric detection of a forward scattered probe laser.
Information gained from the measurement can be used to achieve
conditional spin squeezing. \label{Figure::Schematic}} \vspace{-3mm}
\end{figure}

Conditional spin-squeezing experiments operate on the principle that
weakly measuring $\Fz$ gradually reduces its uncertainty below that
of the initially-prepared spin state. A typical apparatus for such
procedures is depicted in Fig.~\ref{Figure::Schematic}. Continuous
measurement of $\Fz$ is implemented by passing a linearly polarized
probe laser through an atomic sample prepared in an initial
(approximate) coherent state \cite{Polzik2004b} by optical pumping
\cite{Happer1972}. Qualitatively speaking, the atoms rotate (via
state-dependent optical activity) the probe polarization by an
amount proportional to $\Fz$ \cite{Jessen2003}. In a quantum
analysis the atoms and optical probe field evolve into an entangled
state \cite{Takahashi1999} as the result of this scattering
interaction. Detection of the scattered probe field then yields
information about $\Fz$ via these correlations
\cite{Kuzmich1998,Thomsen2002,Thomsen2002b,Thomsen2002c,Silberfarb2003}.

The interpretation of existing squeezing experiments has (at least
implicitly) assumed that polarimetric detection of the forward
scattered probe laser yields a detector photocurrent,
\begin{equation} \label{Equation::BasicPhotocurrent}
    y_t dt = \sqrt{M} \Fz dt + dW_t.
\end{equation}
Here $M$ is a constant (known as the \textit{measurement strength})
that describes the rate at which photodetection provides information
about $\Fz$. The $dW_t$ are noise increments which
exhibit Gaussian white noise statistics $\mathbbm{E}[dW_t] = 0$ and
$dW_t^2=dt$ \cite{vanHandel2005}.

The measurement strength, $M$, is the key parameter for predicting
the degree of squeezing that can be achieved as a result of the
measurement. It is thus important to determine $M$ in terms of
physical properties of the atomic sample and optical probe. While
the form of Eq.\ (\ref{Equation::BasicPhotocurrent}) has been
derived in previous analyses
\cite{vanHandel2005, Kuzmich1998,Thomsen2002,Thomsen2002b,Thomsen2002c,Silberfarb2003},
this has generally been done relative to a simplifying assumption
that the atoms behave qualitatively as spin-$\frac{1}{2}$ particles.
But measurement-induced spin squeezing experiments have utilized
Alkali atoms with higher spin
\cite{Kuzmich2000,Polzik2004,Geremia2004a}, and recent data show
that the deviation from spin-$\frac{1}{2}$ behavior can be
significant \cite{Jessen2004}. While nonlinearities in the
atom-probe scattering process are not always bad (proposals for
capitalizing on these effects for quantum state tomography are being
explored \cite{Silberfarb2005}), they do raise complications for
spin-squeezing experiments by invalidating the form of
Eq.~(\ref{Equation::BasicPhotocurrent}). 

We find that the photocurrent in
Eq.~(\ref{Equation::BasicPhotocurrent}) can be recovered even for
higher-spin atoms by suppressing tensor scattering interactions via
a properly chosen experimental geometry. Using standard techniques
\cite{Happer1967,Deutsch1998} to address the atom-probe scattering
physics, combined with a semiclassical treatment of the atomic
magnetization vector, we derive an expression for the measurement
strength, $M$, in terms of characteristic experimental parameters.
This allows us to obtain an expression for the photocurrent in terms
of the duration of the measurement and the properties of the atomic
system and the probe laser. We observe close agreement between our
scattering model and data obtained using an apparatus of the type in
Fig.\ \ref{Figure::Schematic}.

Finally, we derive an expression for the signal-to-noise ratio of the measurement photocurrent which can be used to calculate the rate of spin-squeezing in experiments of the form shown in Fig.~\ref{Figure::Schematic}. The results we obtain are valid in
the short-measurement limit in which atomic decoherence due to
scattering probe light in unobserved (non-paraxial) electromagnetic
field modes \cite{Carmichael2005} can be safely ignored. Current
spin-squeezing experiments all fall into this regime where the
degree of quantum uncertainty reduction is small compared to the
Heisenberg spin-squeezing limit \cite{Kitagawa1993}.

\section{Continuous Measurement and the Multilevel Atom-Probe
Hamiltonian} \label{Section::Scattering}

We begin by considering the experiment depicted in the schematic of Fig.~\ref{Figure::Schematic}.  An off-resonant linearly polarized probe beam is sent through a cloud of cold spin-polarized atoms.  The forward scattered polarization state of the light is then detected using a polarimeter, consisting of polarization shifting waveplates, a polarizing beam splitter, and two detectors..

In general, it is a rather complicated problem to predict the output polarization state of the probe beam after it has passed through the spatially extended atom cloud.  We can simplify the problem to one dimension by assuming that the beam is predominantly forward scattered due to the coherent re-radiation from a large number of atoms.  This approximation can be extracted from a full three-dimensional model of the diffraction as considered in references \cite{Polzik2004c, Duan2002, Molmer2002}.  Under this assumption, we only consider paraxial modes of the beam.  Neglecting non-paraxial modes prevents us from computing the decoherence rate of the atomic magnetization, but it does not limit our ability to analyze the dynamics in the small-decoherence (short measurement time) limit.  

Even in the one-dimensional problem, the depth of the atomic cloud along the probe direction introduces further complications.  To simplify further, we assume that the overall optical density of the cloud is small enough that the total rotation of the optical polarization state due to the atoms is small.  This allows us to neglect \textit{propagation effects} by which the atoms in the front edge of the cloud would see a substantially different input state than the back edge of the cloud.  These effects lead to complicated multi-mode dynamics which are considered (along with the tensor polarizability) in \cite{Polzik2004}.

Under these approximations, we approach the simplified scattering problem as follows.  The probe beam consists of two orthogonal polarizations and, for each polarization component, the continuous beam is divided into a series of distinct spatial traveling-wave modes, each with a length equal to the depth of the atomic cloud, $L$.  Thus each atom interacts with a pair of orthogonal polarization modes with the same spatial profile for a length of time $\delta t = L/c$.  This approach clearly avoids propagation effects by allowing all of the atoms to interact with the same modes simultaneously.  After the two polarization modes corresponding to one time-slice have interacted with the cloud for the discrete time $\delta t$ , those two modes are detected with the polarimeter, a new time-slice begins to interact with the cloud, and the process is repeated, leading to a continuous measurement.  More detailed approaches to continuous measurement with discrete modes can be found in references \cite{Caves1987, Silberfarb2003}.

Now we define the electric-field and polarization operators associated with each of these optical modes before considering the Hamiltonian interaction of probe beam with a single multilevel atom.  This procedure is discussed with more detail in Appendix \ref{Section::TensorApp}.

\subsection{Probe Field Polarization States}

For each traveling-wave spatial mode, we consider the field operators
\begin{equation} \label{Equation::ENegative}
  \ENegative = \sqrt{\hbar g}  \left[ \aMinusDagger \eMinusStar + \aPlusDagger \ePlusStar \right]
\end{equation}
and
\begin{equation} \label{Equation::EPositive}
  \EPositive = \sqrt{\hbar g} \left[ \aMinus \eMinus + \aPlus \ePlus \right],
\end{equation}
where $\aMinusDagger$ and $\aMinus$ are Heisenberg-picture creation and annihilation operators for the $z$-axis propagating mode with left circular polarization and $\aPlusDagger$ and $\aPlus$ are the creation and annihilation operators for right circular polarization.  Each field operator implicitly refers to a single traveling mode as discussed above, and we neglect to provide indices for the modes unless they are required for clarity.  The coefficient $g=\omega_0/(2\epsilon_0 V)$ is a form factor, $V$ will be taken to be the volume of the atomic cloud, and $\eMinus$ and $\ePlus$ are the (complex) spherical basis vectors for left and right helicity.\\

In the expansion of the polarizability Hamiltonian we get terms which can be recast as Schwinger boson operators
\begin{eqnarray} \label{Equation::Stokes}
  \sZero  &=&  \frac{1}{2} \left( \aPlusDagger\aPlus + \aMinusDagger\aMinus \right) \\
  \sX & = & \frac{1}{2} \left( \aPlusDagger\aMinus + \aMinusDagger\aPlus \right) \nonumber\\
  &=& \frac{1}{2}\left(\aYDagger \aY - \aXDagger \aX\right)\nonumber\\
  \sY & = & \frac{i}{2} \left(  \aMinusDagger\aPlus  - \aPlusDagger\aMinus \right) \nonumber\\
  &=& \frac{1}{2}\left( \aYpDagger \aYp - \aXpDagger \aXp \right)\nonumber\\
  \sZ & = & \frac{1}{2} \left( \aPlusDagger\aPlus - \aMinusDagger\aMinus \right) \nonumber
\end{eqnarray}
These operators obey the usual angular momentum commutation relations and the components form a basis for the Stokes vector which is used to represent the polarization state of the light.  The quantity $\sZero$ is proportional to the number of photons interacting with the atomic system in one time increment.  On any given measurement, the quantity $\sZero$ and a single component of the Stokes vector representing the polarization state (e.g., $\sX$) can be measured with an appropriate selection of polarization rotating waveplates situated after the atoms and prior to the polarizing beam-splitter.  In the usual configuration (of Fig.~(\ref{Figure::Schematic})), $\sX$ is measured without any waveplates, $\sY$ is measured with a half-waveplate that rotates the linear polarization by $45$-degrees, and $\sZ$ is measured by adding a quarter-waveplate that completely circularizes linear polarized light.

In the case where a full quantum mechanical description is used, this choice of basis will change the nature of the information gained from the measurement which is then used to update the conditional collective quantum state describing the atoms.  In other words, the choice of basis will lead to a different \textit{unravelling} of the conditional dynamics.

%\begin{eqnarray}
%\left[ \hat{S}_x , \hat{S}_y \right] &=& i \hat{S}_z \nonumber\\
%\left[ \hat{S}_z , \hat{S}_x \right] &=& i \hat{S}_y \nonumber\\
%\left[ \hat{S}_y , \hat{S}_z \right] &=& i \hat{S}_x
%\end{eqnarray}

\subsection{Scattering Hamiltonian}

We now introduce the polarizability Hamiltonian that determines the joint
evolution of the single-atom spin and the polarization of the traveling-wave optical mode. Subsequently, we summarize the results from Appendix \ref{Section::TensorApp} where we derive a more convenient and intuitive way of representing the irreducible components of the
Hamiltonian in terms of atomic spin operators instead of dipole operators.

For a field which is off-resonant to the transition of interest, the
usual dipole Hamiltonian can be approximated and recast into a
polarizability form.  This can be derived, for example, by using
adiabatic elimination under the assumption that the off-resonant
field only weakly populates the excited states.  The polarizability
Hamiltonian  \cite{Deutsch1998,CohenTannoudji,Happer1972} is then
expressed as
\begin{equation} \label{Equation::PolarizabilityHamiltonian}
  \hInt =   \sum_{f,f'} \ENegative \cdot
     \frac{\Pf \dVec \Pfp \dVecDagger \Pf} {\hbar\Delta_{f,f'}}  \cdot\EPositive.
\end{equation}
where we omit indices identifying the particular atom and spatial optical mode being considered. 
This definition consists of several terms which are also defined in Appendix \ref{Section::TensorApp}.  The negative and positive frequency probe field operators, $\ENegative$ and $\EPositive$, describe the creation and annihilation of photons in the contributing  probe modes.  The atomic operators $\dVecDagger$ and $\dVec$ are the vector dipole raising and lowering operators.  The ground and excited state angular momentum numbers are give by $f$ and $f'$ respectively.  The probe detuning, $\Delta_{f,f'}=\omega-\omega_{f, f'}$, is defined as the difference between the probe frequency $\omega$ and a particular atomic resonance frequency.  For the purposes of this paper we consider all of the population to remain in one ground state manifold ($f=4$ for Cesium, ignoring $f=3$) and sum only over the excited states ($f'=2,3,4,5$). The operators $\Pf$ and $\Pfp$ are projectors onto the ground and excited states respectively.

This Hamiltonian has a satisfying physical interpretation as a scattering interaction: the atom is first brought from its ground state to a virtual excited state via the raising operator, $\dVecDagger$, by annihilating a photon from the probe field through $\EPositive$.  Then, the temporarily excited atom returns to a (potentially different) ground state by emitting a photon into a (potentially different) scattered probe mode via $\dVec$ and $\ENegative$.

The central operator in the scattering Hamiltonian,
\begin{equation}
    \alphaTensor_{f,f'} = \Pf \dVec \Pfp \dVecDagger \Pf,
\end{equation}
commonly called the \textit{atomic polarizability tensor}, is a dyad involving vector operators \cite{Happer1967}.  Thus $\alphaTensor_{f,f'}$ is a rank-2 spherical tensor that can be decomposed into irreducible components,
\begin{equation}
    \alphaTensor_{f,f'} = \alphaTensorZero_{f,f'}  + \alphaTensorOne_{f,f'}  + \alphaTensorTwo_{f,f'}  .
\end{equation}
The scattering Hamiltonian similarly decomposes into irreducible spherical tensor operators,
\begin{equation}\label{Equation::HDecomp}
    \hInt = \hIntZero + \hIntOne + \hIntTwo
\end{equation}
where
\begin{equation} \label{Equation::HComponent}
\hInt^{(j)}=\sum_{f,f'} \ENegative \cdot
     \frac{\alphaTensorJ_{f,f'}} {\hbar\Delta_{f,f'}}  \cdot\EPositive.
\end{equation}
The $\hIntZero$ is a scalar contribution, $\hIntOne$ transforms as a vector, and $\hIntTwo$ transforms as a rank-2 symmetric tensor in the group representation theory of $SO(3)$.   Were the atomic system composed of spin-$\frac{1}{2}$ particles, it would be possible to neglect the rank-2 Hamiltonian \cite{Deutsch1998} (as will become explicit), however, we can not do so for higher-spin Alkali atoms \cite{Jessen2004,Polzik2004,Mueller2004}.

The full Hamiltonian for the collective atomic spin resulting from $N$ atoms is obtained by taking the symmetric sum of these single particle operators.

\subsection{Hamiltonian Decomposition}

Now we recast the single atom Hamiltonian of Eqs.\ (\ref{Equation::HDecomp}, \ref{Equation::HComponent}) into irreducible terms involving only atomic spin operators $\hat{f}_i$ and probe polarization operators $\hat{S}_i$ then discuss each component in physical terms.  The derivation of these expressions is sketched in Appendix \ref{Section::TensorApp}.

\subsubsection{The Scalar Hamiltonian}

The scalar scattering Hamiltonian, $\hIntZero$, can be represented as a product of operators on the separate atomic and probe field Hilbert spaces.  This is accomplished by combining the expressions for the field mode operators, Eqs.\ (\ref{Equation::ENegative}) and (\ref{Equation::EPositive}), with the rank-0 irreducible component of the atomic polarizability tensor.  Evaluating this Hamiltonian using the form of the rank-0 atomic polarizability derived in Appendix \ref{Section::TensorApp} leads to the scalar scattering Hamiltonian,
\begin{eqnarray}
  \hIntZero & = &  g \sum_{f^\prime} \frac{\alphaZero}{\Delta_{f,f^\prime}}
        \frac{2}{3}\sZero \fIdent . \label{Equation::ScalarHamiltonian}
\end{eqnarray}
where the constants $\alphaZero$, defined in equation (\ref{Equation::AlphaZero}) of the appendix, are related to the transition dipole matrix elements for the atomic hyperfine transitions.

This rank-0 Hamiltonian couples the atomic identity operator $\fIdent$ to the field mode number operator and can be interpreted as an atomic state-independent light shift.  It therefore affects both polarization modes of the probe field in an equivalent manner and will not influence the measurement process since it does not provide any state-dependent information.  However, this Hamiltonian would be important if the measurement was meant to distinguish between populations across hyperfine states (e.g. $f=3$ and $f=4$ using homodyne detection) instead of across the sub-level populations within one hyperfine state (using polarimetry, as discussed here).  This term is also of importance if the Hamiltonian is being considered as a spatially dependent potential for the atoms (e.g., in an optical lattice).

\subsubsection{The Vector Hamiltonian}

The vector contribution to the atom-probe scattering Hamiltonian, can be evaluated in a similar manner using expressions for the rank-1 polarizability derived in the appendix,
\begin{eqnarray}\label{Equation::VectorHamiltonian}
  \hIntOne & = &  g \sum_{f^\prime} \frac{\alphaOne}{\Delta_{f,f^\prime}}
     \sZ \fZ.
\end{eqnarray}
Here, the vector polarizability constant, $\alphaOne$, is given by Eq.\ (\ref{Equation::AlphaOne}), $\fZ$ is the $z$-component of the (single-particle) atomic spin angular momentum.

The rank-1 Hamiltonian can be interpreted as causing a differential phase shift on the two circular polarization modes by an amount that is proportional to the $z$-component of the atomic angular momentum.  Thus the vector Hamiltonian leads to optical activity in the atomic sample and produces the familiar Faraday rotation effect often used to address continuous measurement of collective spin \cite{Kuzmich1998,Thomsen2002b,Thomsen2002c,Silberfarb2003,Jessen2003}.  \\

\subsubsection{The Tensor Hamiltonian}

Finally the tensor Hamiltonian, can be evaluated using expressions for the rank-2 polarizability derived in the appendix to give,
\begin{eqnarray}\label{Equation::TensorHamiltonian}
  \hIntTwo & = &  g \sum_{f^\prime}  \frac{\alphaTwo}{\Delta_{f,f^\prime}} \left(
              \sX \left( \fXSquared - \fYSquared \right)\phantom{\frac{2\sZero}{\sqrt{6}}} \right. \\
            & & \quad\quad  + \sY \left( \fX\fY + \fY\fX\right) \nonumber \\
            & & \quad\quad  + \left. \sZero\left( 3 \fZSquared - f(f+1) \fIdent \right)/3\right) \nonumber .
\end{eqnarray}
Here, the tensor polarizability constant, $\alphaTwo$, is given by Eq.\ (\ref{Equation::AlphaTwo}).

The rank-2 Hamiltonian couples spin coordinates to the elliptical components of the probe laser field and produces a second-order light shift proportional to the atomic quadrupole moment.  These terms vanish for $f=1/2$ (as can be seen by evaluating the operators within the parentheses above) but are non-zero for any higher spin number.  For a linearly polarized input beam, the tensor term leads to an elliptically polarized scattered probe field \cite{Happer1967,Polzik2004}.  The rank-2 interaction potentially limits the validity of any analysis of a continuous measurement of collective atomic spin in Alkali atoms based on the qualitative behavior of spin-$\frac{1}{2}$ particles.  

\subsection{Semiclassical Evolution of Probe State}

We can greatly simplify the dynamics by eliminating atomic evolution due to the probe beam and only considering the evolution of the probe beam due to the atomic state.  Under this semiclassical approximation, we replace all atomic operators with their expectation values with respect to an assumed fixed spin state.  (This is the opposite of the semiclassical situation often considered in atom-light interactions where the atomic system is considered quantum mechanically while the optical beam is made classical.)  For a large ensemble of atoms and small interaction times, fixing the atomic state will accurately reproduce the mean behavior of the measured photocurrent corresponding to one of the Stokes vector components.  This is confirmed experimentally in the next section, where the atomic state is fixed and adiabatically positioned with a magnetic holding field.  The holding field serves to both position the atomic state and protect it from the influence of the probe light, such that the analysis of this section remains valid even for long interaction times or large optical depth clouds.  Ultimately, however, probe induced decoherence will dominate all interactions.  In the final section, we then reconsider the full analysis including the atomic quantum noise (related to spin squeezing) for a particular alignment of the collective spin state.

\begin{figure}[t]
\begin{center}
\includegraphics{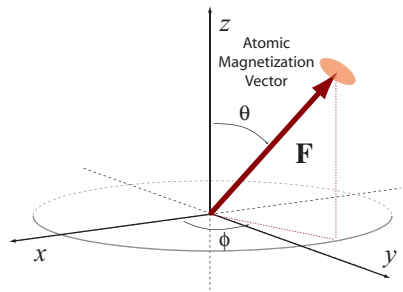}
\end{center}
\caption{Definition of the spherical coordinate angles used to describe the orientation of the collective atomic magnetization vector, $\mathbf{F}$, relative to the fixed laboratory cartesian coordinate system.  The polarization vector of the input probe light resides in the $xy$-plane and forms an angle, $\phi_p$, with respect to the laboratory $x$-axis.  \label{Figure::AngleDefinitions}}
\end{figure}

We approximate the $N$-atom Hamiltonian, $\hInt_N$, by replacing the single-atom operators with their expectation value taken with respect to an optically pumped spin pointing with direction $\theta$ and $\varphi$ given in spherical coordinates (Fig.\ \ref{Figure::AngleDefinitions}).  In other words, for an individual atom operator $\hat{O}_f$, we take
\begin{equation}
\hat{O}_f \rightarrow \langle \hat{O}_f \rangle = \langle \Psi(\theta,\varphi) | \hat{O}_f | \Psi(\theta,\varphi) \rangle
\end{equation}
where $|\Psi(\theta,\varphi)\rangle = \exp[-i \hat{f}_z \varphi]\exp[-i \hat{f}_y \theta] |f,f\rangle_z$.

The relevant operators from the Hamiltonian decomposition are given by
\begin{eqnarray}
     \langle \fZ \rangle &=& f \cos\theta \label{Equation::FzDefinition} \\
     \langle \fXSquared-\fYSquared \rangle &=& f(f-1/2)\sin^2\theta\cos 2 \varphi\\
     \langle \fX\fY-\fY\fX \rangle &=& f(f-1/2)\sin^2\theta\sin 2 \varphi
\end{eqnarray}
Within the semiclassical approximation, we obtain an effective scattering interaction Hamiltonian that only involves operators on the probe field Hilbert space.  Ignoring all terms proportional to $\sZero$ (because it commutes with each term of the semiclassical Hamiltonian) we have
\begin{eqnarray}
\tilde{H} &=& \tilde{H}_N^{(1)} + \tilde{H}_N^{(2)}\nonumber\\
&=& (\gamma_x \sX + \gamma_y \sY + \gamma_z \sZ)\frac{\hbar}{\delta t}
\end{eqnarray}
which leads to a rotation of the Stokes vector $\mathbf{\hat{S}}$ about a vector $\vec{\gamma}=[\gamma_x, \gamma_y, \gamma_z]$ according to the evolution operator

\begin{eqnarray} \label{Equation::SemiclassicalPropagator}
    \UtTilde & = &  \exp\left[ -i \tilde{H} \frac{\delta t}{\hbar}\right]\nonumber\\
    &=& \exp \left[ -i ( \gamma_\mathrm{x} \sX + \gamma_\mathrm{y} \sY + \gamma_\mathrm{z} \sZ)\right]
\end{eqnarray}
where $\delta t=L/c$ is the interaction (transit) time of the discrete spatial modes of the probe beam across the atomic cloud of length $L$. The rotation vector $\vec{\gamma}$ is defined by

\begin{eqnarray}
    \gamma_x & = & \gamma_0 f(f-1/2) \sin^2\theta\cos 2 \varphi \sum_{f^\prime}
                    \frac{\alpha^{(2)}_{f,f^\prime}}{\alpha_0 \Delta_{f,f^\prime}}  \label{Equation::gammaX} \\
    \gamma_y & = &  \gamma_0 f(f-1/2) \sin^2\theta\sin 2 \varphi \sum_{f^\prime}
                    \frac{\alpha^{(2)}_{f,f^\prime}}{\alpha_0 \Delta_{f,f^\prime}}  \label{Equation::gammaY}\\
    \gamma_z & = &  \gamma_0 f \cos \theta \sum_{f^\prime}
                    \frac{\alpha^{(1)}_{f,f^\prime}}{\alpha_0 \Delta_{f,f^\prime}} \label{Equation::gammaZ}
\end{eqnarray}
Here we have normalized by the state-independent polarizability constant (see Appendix \ref{Section::TensorApp})
\begin{eqnarray} \label{Equation::AlphaConstant}
    \alpha_0  &=& \frac{3 \epsilon_0 \hbar \Gamma \lambda_0^3}{8 \pi^2} \\
    &=&   \left |  \langle j || \dVec || j' \rangle \right |^2 \frac{(2 j+1)}{(2j^\prime+1)} \nonumber
\end{eqnarray}
such that $\alpha^{(j)}_{f,f^\prime}/\alpha_0$ is dimensionless.  The rotation strength is represented by
\begin{eqnarray}
    \gamma_0 &=& \frac{N g \delta t \alpha_0}{\hbar}\nonumber
\end{eqnarray}
where we have used the field coefficient $g = \omega_0/(2 \epsilon_0 V)$, the atomic resonance frequency  $\omega_0$, and the interaction volume (the volume of the atomic sample) $V=A L$.

From an experimental standpoint, it is useful to note that $\gamma_0$ is directly related to the on-resonance optical depth $\OD$ of the atomic sample and the decay rate $\Gamma$ via,
\begin{equation}
    \gamma_0 = \left( \frac{\Gamma}{4} \right) \OD
\end{equation}
where
\begin{equation}
    \OD = N \frac{\sigma_0}{A}, \quad \sigma_0 = \frac{3 \lambda_0^2}{2 \pi} .
\end{equation}
The quantity, $\sigma_0$, is the resonant atomic scattering cross section and $A = \pi r^2$ is the cross-sectional area of the atomic sample.

In Appendix \ref{Section::RotateApp} , the equations for a general rotation of $\mathbf{\hat{S}}$ about $\vec{\gamma}$ are given.  Here we specialize to the case where the input beam is linearly polarized in the $\mathrm{x}$-direction such that $\langle \hat{S}_y \rangle = \langle \hat{S}_z \rangle = 0$.  The output expectation values are then given by
\begin{eqnarray}\label{Equation::RotateS}
\langle \hat{S}'_x \rangle &=& \langle \hat{S}_x \rangle\left(\cos\gamma+\frac{\gamma_x^2}{\gamma^2}(1-\cos\gamma) \right)\\
\langle \hat{S}'_y \rangle &=& \langle \hat{S}_x \rangle \left(- \frac{\gamma_z}{\gamma}\sin\gamma+\frac{\gamma_y\gamma_x}{\gamma^2}(1-\cos\gamma) \right)\nonumber\\
\langle \hat{S}'_z \rangle &=& \langle \hat{S}_x \rangle \left(\frac{\gamma_y}{\gamma}\sin\gamma+\frac{\gamma_z\gamma_x}{\gamma^2}(1-\cos\gamma) \right)\nonumber
\end{eqnarray}
Taking the total rotation angle small ($\gamma \ll 1$) this becomes (to second order in $\gamma$)
\begin{eqnarray}
\langle \hat{S}'_x \rangle &\approx& \langle \hat{S}_x \rangle \left(1-\gamma_z^2/2-\gamma_y^2/2\right)\\
\langle \hat{S}'_y \rangle &\approx& \langle \hat{S}_x \rangle \left(-\gamma_z+\frac{\gamma_y\gamma_x}{2} \right)\label{Equation::sYPrime}\\
\langle \hat{S}'_z \rangle &\approx& \langle \hat{S}_x \rangle \left(\gamma_y+\frac{\gamma_z\gamma_x}{2} \right)\label{Equation::sZPrime}
\end{eqnarray}

In this semiclassical approximation, we have completely neglected any evolution of the atomic state due to the probe beam.  We demonstrate in the next section that the above model agrees well with experimental data when the spin state is fixed with a magnetic holding field.

\begin{figure*}
\begin{center}
\includegraphics{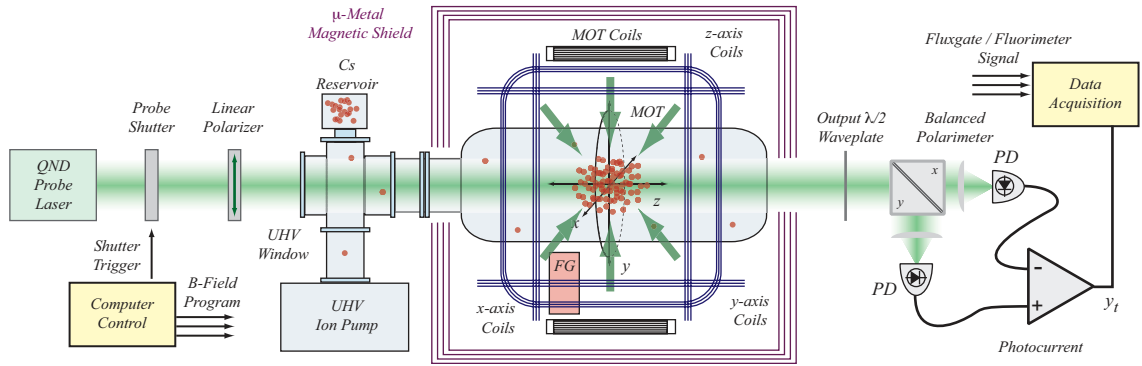}
\end{center}
\vspace{-3mm}
\caption{Schematic of our experimental apparatus in which collective spin angular momentum of a cloud of laser cooled Cs atoms is measured by polarimetric detection of a scattered off-resonant probe laser.  Ambient magnetic field fluctuations are supressed by magnetic shielding and can be monitored with a fluxgate magnetometer (FG) situated nearby the atomic sample.  Components not shown include the optical pumping laser (aligned along the laboratory $x$-axis) and external trim coils used to zero ambient magnetic fields and their first order gradients.  \label{Figure::FullSchematic}}
\end{figure*}

\section{Experimental Results} \label{Section::Validation}

In this section, we show that the model described above is consistent with representative data collected from our experiment with laser cooled Cs atoms and balanced polarimetric detection of a forward-scattered, off-resonant probe laser field.

\subsection{Experimental Apparatus} \label{Section::Experiment}

Figure \ref{Figure::FullSchematic} provides a schematic of the major components of the experimental apparatus.  Our single-particle Alkali atom spin system is the 6$^2$S$_{1/2}$(f=4) ground state hyperfine manifold in $^{133}$Cs with $4 \hbar$ of intrinsic angular momentum due to a combination of the $i=7/2$ nuclear spin and the $s=1/2$ spin of an unpaired 6s valence electron.  We obtain cold atom samples from a $10^{-9}$ Torr background Cs vapor using standard laser cooling and trapping techniques by collecting more than $10^9$ atoms in a magneto-optic trap (MOT).  Trapping beams are derived from a 150 mW injection-locked diode laser tuned (11-15) MHz red of the Cs 6$^2$S$_{1/2}$(f=4)$\rightarrow$6$^{2}$P$_{3/2}$(f$^\prime$=5) cycling transition.  Each 35 mW trapping beam has an approximately constant intensity profile and a 2.5 cm diameter. A 10 mW repump laser tuned to the 6$^2$S$_{1/2}$(f=3)$\rightarrow$6$^{2}$P$_{3/2}$(f$^\prime$=4) transition is used to prevent atomic population from decaying out of the trapping cycling transition.

Following the atom collection phase, the sample is sub-Doppler cooled \cite{Gould1988} to a temperature of $T\sim 10$ $\mu$K and the initial $x$-polarized spin state is prepared with a circularly polarized 100 $\mu$W optical pumping beam (pulsed for 2-4 ms) propagating along the $x$-axis and tuned to the (f=4)$\rightarrow$(f$^\prime$=4) hyperfine transition.  A 100 mG magnetic holding field is applied along the laboratory $x$-axis to define the optical pumping direction.

Continuous measurement of the polarized atomic ensemble is implemented with a nearly quantum shotnoise-limited probe laser that can be detuned from the 6$^2$S$_{1/2}$(f=4)$\rightarrow$6$^2$P$_{3/2}$(f$^\prime$=5) Cs transition over a range $\Delta=\pm 1.4$ GHz. The probe beam is linearly polarized by a high extinction Glan-Thompson prism prior to passing through the cold atom cloud, and the orientation of the linear polarization vector with respect to the laboratory coordinate system may be rotated via an input half-waveplate.  The scattered probe field is detected with a polarimeter constructed from a Glan-Thompson polarizing beam splitter and a DC-balanced photodetector with $>$1 MHz measurement bandwidth.

\begin{figure*}
\begin{center}
\includegraphics{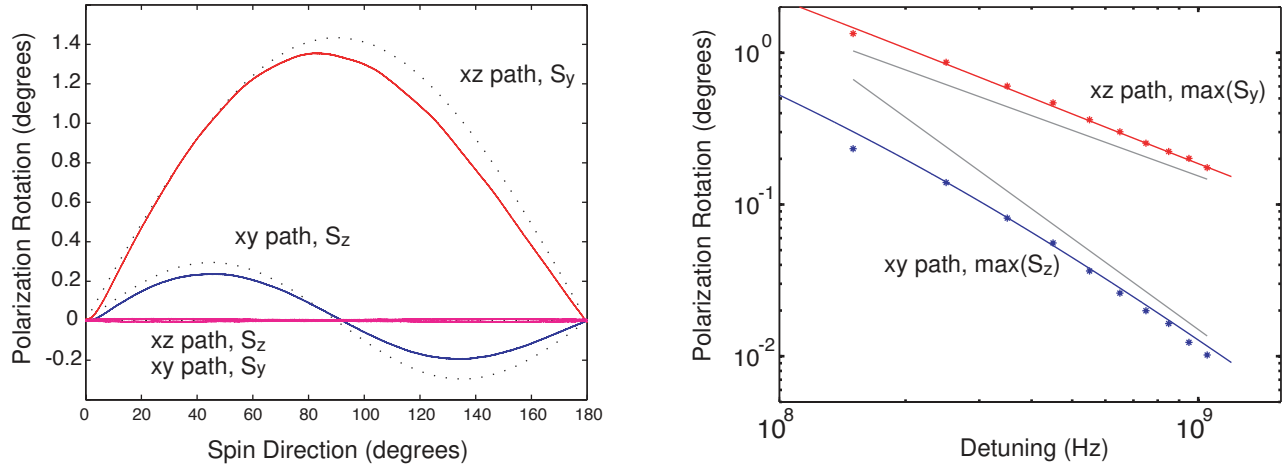}
\end{center}
\vspace{-5mm}
\caption{Comparison of our model of continuous measurement with photocurrents obtained from from the experiment with $N=1 \times 10^9$ Cs atoms in an $r=4$ mm spherical trap and a $P=10$ $\mu$W probe field blue-detuned from the (f=4)$\rightarrow$(f$^\prime$=5) $D_2$ hyperfine transition ($\lambda_0=$ 852 nm). Each trajectory is averaged 10 times.  (A) For an input probe beam with x-polarization and a detuning of 150 MHz, $\hat{S}_{\mathrm{y}}$ and $\hat{S}_{\mathrm{z}}$ were measured for both the $xz$ and $xy$ trajectories (described in the text) resulting in the solid curves.  All trajectory times are $\tau=2$ ms, during which we observe some atomic decoherence which causes the prediction (dotted curves) to stray from the data. (B)  As a function of probe detuning, we plot the peak of the $\hat{S}_{\mathrm{y}}$ measurement (for the $xz$ trajectory) which depends only on rank-1 terms and the peak of the $\hat{S}_{\mathrm{z}}$ measurement (for the $xy$ trajectory) which depends only on rank-2 terms. The predicted behavior (solid curves) shows good agreement with the data out to large detunings where the curves asymptote to the $1/\Delta$ and $1/\Delta^2$ lines provided to guide the eye.}
\label{Figure::StokesData}
\end{figure*}

A computer controls the experiment timing and records the polarimeter output as well as diagnostic information including background magnetic field fluctuations (measured with a flux-gate magnetometer) and atom number (measured by fluorescence imaging).  The computer enables/disables the measurement by controlling a shutter on the probe laser, constructed from a switched acousto-optic modulator, with 100 ns resolution.  Magnetic fields with magnitudes up to $\sim 0.5$ G can be applied in arbitrary (time-dependent) directions by driving 3 pairs of computer-controlled Helmholtz coils, oriented along the laboratory $x$-, $y$-, and $z$-axes, with a bandwidth of $\sim 1$ MHz.

Background magnetic field fluctuations are suppressed through a combination of passive $\mu$-metal shielding and field cancellation via external trim coils.  Each atom preparation (trapping, cooling and optical pumping) and measurement cycle is synchronized with respect to the 60-Hz building power lines to suppress the effects of induced magnetic fields.  Slow magnetic drift due to natural and anthropogenic sources are cancelled by adjusting the external trim coils based on the output of the fluxgate magnetometer.

\subsection{Verification of the Probe Scattering Model} \label{Section::StokesVerification}

Our model of the scattered probe polarization as a function of the orientation of the atomic magnetization vector was compared against experiment by observing the polarimeter photocurrent as the orientation of the atomic polarization was varied according to different specified paths in the laboratory coordinate system.  This was accomplished as follows.  An $x$-polarized cold atom sample was prepared according to the description above and an $x$-axis magnetic holding field of 100 mG was applied.  At this point, the probe shutter was opened and the balanced polarimeter photocurrent was monitored while the orientation of the magnetic holding field was varied according to the specified path.  The rate of change of the holding field orientation was chosen to be slow (ms) compared to the atomic Larmor precession frequency (hundreds of kHz) such that the atomic magnetization vector adiabatically followed the path traced by the holding field.  Furthermore the holding field was large enough to dominate the probe light induced dynamics at short times, but not so strong as to shift the levels significantly compared to the detuning.  

With a strong enough holding field, the spin state (and hence the semiclassical rotation vector $\vec{\gamma}$) will be fixed across the spatially extended cloud.  Because rotations of the Stokes vector about the same vector will commute, the semiclassical analysis of the previous section will be valid for even large optical depth samples where the total optical polarization rotation is significant.

This process was performed for two different adiabatic paths on the atomic Bloch sphere:
\begin{itemize}

\item \textbf{\textit{xz}-Plane Rotation}: the atomic magnetization follows a path beginning along the $x$-axis and rotates around the $y$-axis:  $\theta=\pi/2\rightarrow -\pi/2$ with fixed $\varphi=0$.

\item \textbf{\textit{xy}-Plane Rotation}: the atomic magnetization follows a path beginning along the $x$-axis and rotates around the $z$-axis: $\varphi=0\rightarrow \pi$ with fixed $\theta=\pi/2$.

\end{itemize}
We chose these two trajectories because they highlight the different contributions from the rank-1 and rank-2 scattering interactions, as seen from Eqs.\ (\ref{Equation::gammaX}, \ref{Equation::gammaY}, \ref{Equation::gammaZ}).  The $xz$-plane trajectory, where $\varphi=0$, virtually eliminates the rank-2 tensor contribution to the photocurrent  leaving nearly ideal Faraday rotation.  Conversely, the $xy$-plane rotation eliminates rank-1 contributions and produces elliptical scattered probe polarizations.

\subsubsection{Measuring the Scattering Probe Stokes Vector}

Fig.\ \ref{Figure::StokesData}A compares the measured polarimeter photocurrents (solid curves) for these two adiabatic trajectories with those predicted by our atom-field scattering model (dotted curves).   The input state was polarized in the $x$ direction and because the total polarization rotation angle induced by the atoms $\gamma$ was small, we measured only the other two components $\sY$ and $\sZ$ with the appropriate arrangement of waveplates prior to the polarimeter.  For measuring $\sY$ a single half-waveplate is placed prior to the polarizing beamsplitter (PBS) to rotate the polarization by $45$ degrees, and for $\sZ$ a quarter-waveplate is used to circularize the initial linearly polarized light.

We now refer to Eq.\ (\ref{Equation::sYPrime}) and Eq.\ (\ref{Equation::sZPrime}) to explain the observed trajectories.  For the $xz$ trajectory, we have $\gamma_y=0$ such that $\sY$ contains a large linear term in $\gamma_z$ but $\sZ$ only contains terms quadratic in $\gamma$.  Thus, for this path, the measurement of $\sY$ leads to the top curve in Fig.\ \ref{Figure::StokesData}A which is proportional to the rank-1 polarizability, while the measurement of $\sZ$ is much smaller and effectively zero.  For the $xy$ trajectory, we have $\gamma_z=0$ such that $\sZ$ contains a large linear term in $\gamma_y$ but $\sY$ only contains terms quadratic in $\gamma$.  Thus, for this path, the measurement of $\sZ$ leads to the middle curve in Fig.\ \ref{Figure::StokesData}A which is proportional to the rank-2 polarizability, while the measurement of $\sY$ is much smaller and effectively zero.  The doubling of frequency between the two dominant curves is a direct consequence of the tensor nature of the rank-2 term.

Note that there is some structure expected in the two curves (quadratic in $\gamma$) which are approximately zero, but these are more polluted by technical noise and do not reveal any essentially new information about the interaction.  For the larger curves (linear in $\gamma$), deviations of the measured photocurrents relative to the predicted values seen in Fig.\ \ref{Figure::StokesData}A result mainly from the fact that the model does not consider the probe-induced damping.

The predicted curves use values for the atom number, trap volume, probe power and detuning consistent with independent characterizations of those parameters.  The atom number and trap volume were obtained from fluorescence detection of the MOT and a CCD image of the atom cloud, and the resulting values, $N=1\times 10^9$ and $r=4$ mm, correspond to an optical depth, $\OD \sim 7$, which is consistent with absorption measurements that we performed.  Given our uncertainty in measuring the number of atoms, it can be inferred that our optical pumping efficiency in these (relatively) optically thin atomic samples is no less than 85\% (but is more likely $>$90\%) \cite{Polzik2004b}.

\subsubsection{Relative Scaling of the Scattering Terms with Probe Detuning}

As further verification of our scattering model, we investigated the scaling of the rank-1 and rank-2 contributions to the polarimeter photocurrent as a function of the probe detuning.   With reference to Fig.~( \ref{Figure::StokesData}A), the magnitude of the vector and tensor scattering interactions were measured from the peak amplitude of the $\sY$ measurement (for the $xz$ plane rotation) and the amplitude of the $\sZ$ measurement (for the $xy$-plane rotation) respectively.  This plot compares these measured signal amplitudes (stars) with those predicted by our scattering model (solid curves) for detunings (with respect to the (f=4)$\rightarrow$(f$^\prime$=5) hyperfine transition) ranging from 150 MHz to 1.05 GHz.

The fact that multiple excited state hyperfine levels participate in the scattering interaction is evident from scalings which are not constant in $\Delta^{-1}$ or $\Delta^{-2}$.   As supported by our full model of the scattering interaction, we observe no qualitative difference in the continuous measurement for probe detunings smaller than the hyperfine splittings.  This suggests that conditional spin-squeezing experiments can be performed with small detunings provided that the probe intensity is weak enough that the small decoherence requirement is satisfied.

\section{Spin-Squeezing with Multilevel Atoms}

Until this point we have considered only the semiclassical evolution of the optical probe beam due to an ensemble of atoms with a fixed atomic spin state.  Now we consider a different experimental scenario appropriate for preparing conditional spin-squeezed states of the atomic ensemble.  As opposed to the previous situation, we remove the adiabatic holding field which makes spin-squeezing impossible as it will cause undesired mixing of the squeezed and anti-squeezed components perpendicular to the mean spin.  Although the holding field may serve to validate the previous semiclassical analysis for longer times by eliminating the probe-induced evolution of the atomic state, this analysis is still valid for small times and weak interactions without a holding field.  Thus our goal is to derive the small time signal-to-noise ratio by deriving the signal strength from the previous section and comparing this to the optical shotnoise.  We then use the signal-to-noise ratio to predict the rate of squeezing in a typical experimental configuration where the tensor terms can be ignored.  

Considering only the relatively low optical density limit, the measurement of $\sY$ will result in Eq.\ (\ref{Equation::sYPrime}).  Now we wish to re-write this equation in the form of Eq.\ (\ref{Equation::BasicPhotocurrent}) including measurement noise.  It is readily shown that all terms not linear in $\Fz$ vanish in Eq.\ (\ref{Equation::sYPrime}) provided that $\theta=\pi/2$ and $\phi =0$.  That is, a pure Faraday rotation Hamiltonian is recovered when the atomic magnetization vector is oriented along the $x$-axis. However, rotating $\F$ in the $xy$-plane results in elliptically polarized scattered probe light, and moving out of this plane results in nonlinear atomic dephasing due to scattering terms which are quadratic in the single-particle spin operators, $\hat{f}_\mathrm{z}$.  These adverse effects are avoided for the experimental geometry where $\F$ is collinear with the $x$-axis.  Fortunately, spin-squeezing experiments are easily operated under such conditions \cite{Geremia2004a}.

Taking the input probe field to be in an $x$-polarized optical coherent state, and considering the small $\gamma$ limit, Equation (\ref{Equation::sYPrime}) leads to a semiclassical photocurrent (with units of optical power) of the form,
\begin{equation} \label{Equation::GoodPhotocurrent}
    y_t = \eta \sqrt{S} \Fz + \sqrt{\eta}\, \zeta_t,
\end{equation}
where we have made the substitution, $\hbar N f \cos\theta \rightarrow \Fz$ (refer to Eq. (\ref{Equation::FzDefinition})), and included the photodetector quantum efficiency, $\eta$.   Note that we have introduced $\zeta_t$ which represents optical shotnoise.  We have also introduced a constant, $S$, the \textit{scattering strength},
\begin{equation}
    S =  \frac{1}{\hbar^2} \left[ I_p \sigma_0 \left(\frac{\Gamma}{4}\right)
        \sum_{f^\prime} \frac{\alphaOne}{\alpha_0 \Delta_{f,f^\prime}} \right]^2,
\end{equation}
that depends up the probe intensity, $I_p = P / A$, determined by the coherent state amplitude, $P= 2 \hbar\omega |\beta|^2$ and cross-sectional area, $A=\pi r^2$ (for a mode-matched probe laser).    It is useful to note that the scattering strength has units of $\mathrm{W}^2/\hbar^2$ (power squared per $\hbar^2$) and characterizes the degree of coupling between the atoms and the probe field;  $\sqrt{S}$ quantifies the polarimeter optical power imbalance per unit spin (as $\Fz$ has units of $\hbar$).

Our expressions are similar to previous results \cite{Thomsen2002, Silberfarb2003, Polzik2003} in that it appears as a Faraday rotation signal.  However, our specific expressions for $\gamma_\mathrm{x}$, $\gamma_\mathrm{y}$ and $\gamma_\mathrm{z}$ account for the detailed hyperfine structure of the atomic excited states, including the fact that the oscillator strengths and signs of the contributions from different participating excited states are not equal, and doing so is required for quantitative agreement between theory and experiment.

To arrive at an expression for the \textit{measurement strength}, $M$, as defined in Eq.\ (\ref{Equation::BasicPhotocurrent}), we must consider the variance, $\Delta \zeta^2$, of the white noise increments $\zeta_t$.  For an optical coherent state \cite{Glauber1963,Mandel1995}, this noise variance is given by the familiar optical shotnoise expression,
\begin{equation}
    \Delta\zeta^2 = \mathbbm{E}[ \zeta_t^2] = 2 \hbar \omega P,
\end{equation}
which has units of $\mathrm{W}^2/\mathrm{Hz}$ (power squared per frequency).   Comparing the semiclassical photocurrent of Eq.\ (\ref{Equation::GoodPhotocurrent}) to the photocurrent of Eq.\ (\ref{Equation::BasicPhotocurrent}), the measurement strength is seen to be given by the ratio
\begin{equation} \label{Equation::MeasurementStrength}
    M = \frac{S}{\Delta \zeta^2} = \frac{1}{2 \hbar^2}\tau_\mathrm{s}^{-1} \left( \frac{\sigma_0}{A}\right) ,
\end{equation}
where we have defined the reciprocal scattering time as
\begin{equation} \label{Equation::ScatteringRate}
    \tau_\mathrm{s}^{-1} = \frac{I \sigma_0}{\hbar \omega}  \left( \frac{\Gamma}{4}
        \sum_{f^\prime} \frac{\alphaOne}{\alpha_0 \Delta_{f,f^\prime}} \right)^2,
\end{equation}
which is essentially the rate that probe photons are scattered by the atomic system.  This expression is similar to that derived in Ref.\ \cite{Jessen2003}.

Now consider a measurement of $\Fz$ by Eq.\ (\ref{Equation::BasicPhotocurrent}).  In the small time limit where probe induced decoherence can be neglected, the full quantum filter describing this measurement is equivalent a classical model in which $\Fz$ is simply a random constant on every trial drawn from a distribution with variance equal to the quantum variance of $\langle \Delta \Fz ^2 \rangle_0$  \cite{Stockton2004}.  Then the generally complicated full quantum filter \cite{vanHandel2005} is equivalent to linear regression, or fitting a constant to the noisy measurement record in real time.  In essence, the optimal filter serves to average away the optical shotnoise to reveal the underlying value of $\Fz$. Under these statistical assumptions, at small times the quantum uncertainty is given by
\begin{equation}
\langle \Delta \Fz ^2 \rangle_\tau = \frac{\langle \Delta \Fz ^2 \rangle_0}{1+\eta \langle \Delta \Fz ^2 \rangle_0 M \tau}.
\end{equation}
This can be shown either with the full quantum filter or by using the equivalent classical model combined with Bayesian estimation (from which a Kalman filter or linear regression can be derived).

\begin{figure}[t]
\vspace{2mm}
\begin{center}
\includegraphics{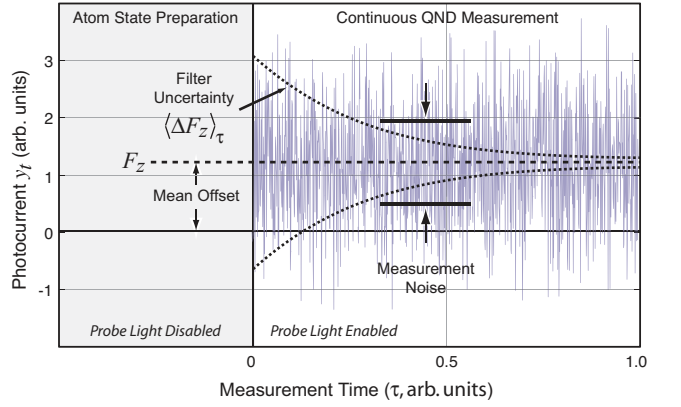}
\end{center}
\vspace{-5mm}
\caption{Simulated photocurrent ($\eta=1$) for a continuous measurement of atomic spin angular momentum via balanced polarimetry.  At the onset of the measurement, $t=0$, the photocurrent assumes a mean offset proportional to the $z$-component of the spin, but this offset is masked by white noise due, in part, to optical shotnoise on the probe laser.  Filtering the photocurrent gradually reduces the uncertainty in the photocurrent offset and produces spin-squeezing. \label{Figure::Photocurrent}}
\end{figure}

These concepts are illustrated by the simulated  measurement trajectory in Fig.\ \ref{Figure::Photocurrent}.  The plot begins with the probe laser turned off, during which all necessary state preparation of the atomic system such as atom trapping, cooling and optical pumping into an $x$-polarized coherent spin state is performed.   Once the probe light is enabled at $t=0$, the photocurrent acquires a mean offset, $\eta \sqrt{S} \Fz$, proportional to the spin measurement outcome, $\Fz$, but this mean value is masked by photocurrent noise.  At short times, the signal is overwhelmed by local statistical fluctuations; however, averaging the photocurrent suppresses the uncertainty in the mean signal by integrating away the white noise, illustrated by the dotted lines in Fig.\ \ref{Figure::Photocurrent}.

If we define the signal to noise ratio as
\begin{equation}
\SNR^2 \equiv \eta \langle \Delta \Fz ^2 \rangle_0 M \tau
\end{equation}
we can then express the degree of squeezing (ignoring decay of the $\Fx$) as
\begin{eqnarray}
W&\equiv&\frac{\langle \Delta \Fz ^2 \rangle_t}{\langle \Delta \Fz ^2 \rangle_0}\nonumber\\
&=& \frac{1}{1+\SNR^2}
\end{eqnarray}
Using  $\langle \Delta \Fz ^2 \rangle_0=\hbar^2 N f /2$, we can express the signal to noise ratio as
\begin{equation}
\SNR^2 = \eta \OD \frac{f}{4} \frac{\tau}{\tau_s}
\end{equation}
To keep this expression valid we must have $\tau\ll\tau_s$, so our only recourse to creating large amounts of squeezing in free space is to increase both the quantum efficiency $\eta$ and the optical depth $\OD$ as much as possible.

\section{Conclusion}

In this work, we have derived the most useful form of the polarizability Hamiltonian describing the realistic measurement of an ensemble of multilevel alkali atoms with an off-resonant probe beam.  We then showed that this model was consistent with experimental observations in the semiclassical limit where the atomic state was adiabatically directed with a strong magnetic field.   We found that an adequate comparison was only possible after including all relevant hyperfine transitions including their relative (non-unit) oscillator strengths in our model of the atomic physics.

We then developed a model for describing conditional spin-squeezing in Alkali atoms.  Detailed investigation of the atom probe scattering physics indicates that it is possible to eliminate unwanted tensor components of the atomic polarizability by adopting a suitable atomic and optical polarization geometry.  This includes the elimination of dephasing due to the quadratic light shift \cite{Jessen2004} without sacrificing a fixed laboratory coordinate system for the measurement.  Moreover, we found that conditional spin-squeezing experiments could be performed at small optical detunings without a substantial change in the form of the photocurrent or filtering approach.

\begin{acknowledgements}
The authors would like to thank Poul Jessen, Ivan Deutsch, Andrew Silberfarb, Dima Budker and especially Ramon van Handel and Andrew Doherty for numerous insightful discussions.  We would also like to thank Sebastian de Echaniz, Jacob Sherson, and Eugene Polzik for pointing out important corrections.  This work was supported by the Caltech MURI Center for Quantum Networks (DAAD19-00-1-0374).  JMG acknowledges support from the Caltech Center for Physics of Information and JKS acknowledges support from a Hertz Fellowship.
\end{acknowledgements}

\bibliography{Paper}

\appendix

\section{The Irreducible Representation of the Polarizability Hamiltonian}\label{Section::TensorApp}

In this appendix, we derive the irreducible components of the polarizability Hamiltonian, Eqs. (\ref{Equation::ScalarHamiltonian}, \ref{Equation::VectorHamiltonian}, \ref{Equation::TensorHamiltonian}), from the less useful form of Eq. (\ref{Equation::PolarizabilityHamiltonian}).  We begin by clarifying notation used for the spherical basis and the spin states of the Alkali atoms.  Then we discuss properties of the polarizability Hamiltonian and the dipole operator before detailing the decomposition and re-formatting of the Hamiltonian into its irreducible form.

\subsection{Spherical basis}

The spherical basis is the preferred basis when dealing with atomic transitions due to its symmetry properties.  The basis is defined by the transformation from Cartesian coordinates
\begin{eqnarray}
\ePlus &=& -(\eX+i \eY)/\sqrt{2}\\
\eMinus &=& (\eX-i \eY)/\sqrt{2}\nonumber\\
\eZero &=& \eZ \nonumber
\end{eqnarray}
Thus elements in the spherical basis have the properties
\begin{eqnarray}
\eStar_q=\e_{-q} (-1)^q\\
\e_q \cdot \eStar_{q'} = \delta_{q, q'}\nonumber
\end{eqnarray}
and for an arbitrary vector $\mathbf{A}$ we have $A_q = \e_q \cdot \mathbf{A}$ so that $\mathbf{A} = \sum_q A_q \eStar_q = \sum_q (-1)^q A_q \e_{-q}$.

\subsection{Alkali spin states}

We represent the internal state of the atom in terms of the (Zeeman degenerate) atomic hyperfine states,  $|f,m\rangle$.  Here $f$ and $f^\prime$ are the total spin quantum numbers for the ground and excited hyperfine levels respectively while $m$ and $m^\prime$ are their projections on the $z$-axis.  That is to say, $|f,m\rangle$ are eigenstates of the total atomic angular momentum,
\begin{equation}
    \hat\mathbf{f} = \hat\mathbf{s}\otimes\hat\mathbbm{1}_{l\otimes i} +
    \hat\mathbbm{1}_s \otimes \hat\mathbf{l}\otimes\mathbbm{1}_{i} +
    \hat\mathbbm{1}_{s\otimes l}\hat\mathbf{i}
\end{equation}
where $\hat\mathbf{s}$, $\hat\mathbf{l}$, and $\hat\mathbf{i}$ are respectively the electron spin, orbital angular momentum, and the nuclear spin.  The quantum numbers, $f$, and $m$, are defined in the usual manner,
\begin{eqnarray}
    \hat\mathbf{f}^2 | f,m \rangle & = & \hbar^2 f(f+1) |f,m\rangle \\
    \hat{f}_\mathrm{z} | f, m \rangle & = & \hbar m |f,m\rangle \nonumber
\end{eqnarray}
We use the notation that $\fPM$ are in the spherical basis
\begin{equation}
\fPM = \mp (\fX \pm i \fY)/\sqrt{2}.
\end{equation}
It will also be useful to define a projector onto the ground state $f$
\begin{equation}
\Pf =  \sum_{m} |  f, m \rangle \langle  f, m |
\end{equation}
and a projector onto the excited state $f'$
\begin{equation}
\Pfp =  \sum_{m'} |  f', m' \rangle \langle  f', m' |.
\end{equation}

\subsection{Hamiltonian approximation}

We begin with the single-particle dipole Hamiltonian $H=-\dVec\cdot \hat{\mathbf{E}}$.  The dipole operator $\dVec=e \hat{\mathbf{r}}_e$ can be split into its raising and lowering components
\begin{eqnarray}
\dVec &=& \dVec^{(-)}+\dVec^{(+)}\\
\dVec^{(-)}&=&\sum_{f, f'} \Pf \dVec \Pfp\nonumber\\
\dVec^{(+)}&=&\sum_{f, f'} \Pfp \dVec \Pf\nonumber
\end{eqnarray}
and the electric field operator can be split into rotating and counter-rotating terms
\begin{eqnarray}
  \hat{\mathbf{E}} &=&  \ENegative + \EPositive \label{Equation::EField}\\
  \ENegative &=& \sqrt{\hbar g}  \left[ \aMinusDagger \eMinusStar + \aPlusDagger \ePlusStar \right]\nonumber\\
  \EPositive &=& \sqrt{\hbar g} \left[ \aMinus \eMinus + \aPlus \ePlus \right]\nonumber
\end{eqnarray}
After using the rotating wave approximation and one of many available perturbation expansion techniques (e.g., adiabatic elimination) we arrive at the familiar polarizability Hamiltonian \cite{CohenTannoudji,Deutsch1998,Happer1972},
\begin{eqnarray}
  \hInt &=&  \sum_{f,f'} \ENegative \cdot
     \frac{\alphaTensor_{f,f'} } {\hbar\Delta_{f,f'}}  \cdot\EPositive
\end{eqnarray}
where the atomic polarizability between a particular ground state ($f$) and excited state ($f'$) is defined as
\begin{eqnarray}
    \alphaTensor_{f,f'} &=& \Pf \dVec \Pfp \dVecDagger \Pf\\
     & = & \sum_{m} \sum_{m^\prime}
  \sum_{m^{\prime\prime}}
  | f,m^{\prime\prime}\rangle \langle f,m^{\prime\prime}| \dVec
  | f^\prime, m^\prime \rangle \\
  && \quad \quad \times \langle
   f^\prime, m^\prime | \dVecDagger
  |  f, m \rangle \langle  f, m | . \nonumber
\end{eqnarray}
This expanded expression involves dipole operator matrix elements of the form, $\langle  f^\prime, m^\prime | \hat{d}_q | f, m \rangle$ where $|f,m\rangle$ is a Zeeman sub-level in the ground-state hyperfine manifold, $|f^\prime,m^\prime\rangle$ is a virtual state in the excited hyperfine manifold, and $q=0,\pm1$ labels the helicity of the electromagnetic field.

The above notation is complete, but for the rest of this appendix we work with only one particular $f, f'$ combination and remove the subscripts with the simplifying notation change
\begin{eqnarray}
\Pf\dVec^{(-)}\Pfp \rightarrow \dVec \\
\Pfp\dVec^{(+)}\Pf \rightarrow \dVecDagger \nonumber\\
\alphaTensor_{f,f'} \rightarrow \alphaTensor \nonumber
\end{eqnarray}
However, when the complete Hamiltonian is considered, the summation over all possible $f, f'$ combinations is re-established.

\subsection{Matrix element decomposition}

In order to work with the above expressions, it is advantageous to simplify the dipole matrix elements as far as possible.  By employing the Wigner-Eckart theorem, the angular dependence of the matrix element, $\langle f^\prime,m^\prime | \dVec | f,m\rangle$ can be factored into the product
of a Clebsch-Gordan coefficient and a reduced matrix element,
\begin{equation}
    \langle f, m | \hat{d}_q | f^\prime, m^\prime \rangle  =  \langle f, m | 1, q ; f^\prime,
    m-q \rangle \langle f || \dVec || f^\prime \rangle  .
\end{equation}
Since the dipole operator acts only on electronic degrees of freedom, it is further possible to
factor out the nuclear spin degrees of freedom via the explicit coupling,
\begin{eqnarray}
  \langle f || \dVec || f^\prime \rangle & = & (-1)^{f^\prime+j+i+1} \sqrt{(2f^\prime+1)(2 j+1)} \nonumber\\
  & &  \,\,\times \left\{ \begin{array}{ccc} 1 & j & j^\prime \\ i & f^\prime & f \end{array} \right\}
  \langle j || \dVec_\mathrm{e} || j^\prime \rangle
\end{eqnarray}
where $i$ is the nuclear spin quantum number, $j$ and $j^\prime$ are the ground and excited state fine structure quantum numbers, and $\dVec_\mathrm{e}$ is the dipole operator with respect to the electronic degrees of freedom.

\subsection{Tensor decomposition}

From Eq. (17-89) of reference \cite{Baym1969} we see that we can form an irreducible tensor, $\hat{Z}^{(j)}_m$, from a linear combination of tensor operators $\hat{U}^{(\kappa)}_q$ and  $\hat{V}^{(\kappa')}_{q'}$ via the definition
\begin{equation}
\hat{Z}^{(j)}_m = \sum_{q, q'} \hat{U}^{(\kappa)}_q \hat{V}^{(\kappa')}_{q'} \langle \kappa, q; \kappa', q' | j, m\rangle
\end{equation}
where $\langle \kappa, q; \kappa', q' | j, m\rangle$ are Clebsch-Gordan coefficients.  This expression can then be inverted using
\begin{equation}
\hat{U}^{(\kappa)}_q \hat{V}^{(\kappa')}_{q'}  = \sum_{j, m} \hat{Z}^{(j)}_m \langle \kappa, q; \kappa', q' | j, m\rangle.
\end{equation}

We now specialize to the case where $\hat{Z}^{(j)}_m=\hat{T}^{(j)}_m$, $\hat{\mathbf{U}}=\dVec$, and $\hat{\mathbf{V}}=\dVecDagger$.  Because we are creating a dyad (with two vectors), we have $\kappa=\kappa'=1$.  Inserting these above gives the definition
\begin{equation}
\hat{T}^{(j)}_m = \sum_{q, q'} \dC_q \dCDagger_{q'} \langle 1, q; 1, q' | j, m\rangle
\end{equation}
and the inverse
\begin{equation}
\dC_q \dCDagger_{q'}  = \sum_{j, m} \hat{T}^{(j)}_m \langle 1, q; 1, q' | j, m\rangle.
\end{equation}

We can use this latter expression to write the polarizability as
\begin{eqnarray}
    \alphaTensor &=& \dVec \dVecDagger \\
    &=&\sum_{q, q'} \eStar_{q} \eStar_{q'} \dC_q \dCDagger_{q'}\\
    &=& \sum_{j, m} \sum_{q, q'} \eStar_{q} \eStar_{q'} \hat{T}^{(j)}_m \langle 1, q; 1, q' | j, m\rangle\\
    &=& \alphaTensorZero \oplus \alphaTensorOne \oplus \alphaTensorTwo
\end{eqnarray}
where
\begin{equation}
\alphaTensorJ = \sum_{m=-j}^{j} \hat{T}^{(j)}_m \sum_{q, q'} \eStar_{q} \eStar_{q'}  \langle 1, q; 1, q' | j, m\rangle\label{Equation::alphaTensorJ}\\
\end{equation}

Filling in these Clebsch-Gordan coefficients explicitly, we get
\begin{eqnarray}\label{Equation::alphaExpand}
\alphaTensorZero &=& \hat{T}^{(0)}_0 \left[ -\frac{1}{\sqrt{3}} \eZeroStar \eZeroStar
+\frac{1}{\sqrt{3}}\ePlusStar\eMinusStar
+\frac{1}{\sqrt{3}}\eMinusStar\ePlusStar \right] \nonumber\\
\alphaTensorOne &=& \hat{T}^{(1)}_0 \left[
\frac{1}{\sqrt{2}} \ePlusStar \eMinusStar
-\frac{1}{\sqrt{2}} \eMinusStar \ePlusStar
\right]\nonumber\\
&&+ \hat{T}^{(1)}_{+1} \left[ -\frac{1}{\sqrt{2}} \eZeroStar \ePlusStar
+\frac{1}{\sqrt{2}} \ePlusStar \eZeroStar \right]\nonumber\\
&&+ \hat{T}^{(1)}_{-1} \left[ \frac{1}{\sqrt{2}} \eZeroStar \eMinusStar
-\frac{1}{\sqrt{2}} \eMinusStar \eZeroStar \right]\nonumber\\
\alphaTensorTwo &=& \hat{T}^{(2)}_{0} \left[ \frac{2}{\sqrt{6}} \eZeroStar \eZeroStar
+\frac{1}{\sqrt{6}} \ePlusStar \eMinusStar
+\frac{1}{\sqrt{6}} \eMinusStar \ePlusStar \right]\nonumber\\
&&+\hat{T}^{(2)}_{+1} \left[ \frac{1}{\sqrt{2}} \eZeroStar \ePlusStar
+\frac{1}{\sqrt{2}} \ePlusStar \eZeroStar \right]\nonumber\\
&&+\hat{T}^{(2)}_{-1} \left[ \frac{1}{\sqrt{2}} \eZeroStar \eMinusStar
+\frac{1}{\sqrt{2}} \eMinusStar \eZeroStar \right]\nonumber\\
&&+\hat{T}^{(2)}_{+2} \left[ \ePlusStar \ePlusStar \right]\nonumber\\
&&+\hat{T}^{(2)}_{-2} \left[ \eMinusStar \eMinusStar \right]
\end{eqnarray}

Furthermore, using the definition of $\hat{T}^{(j)}_m$ and filling in the Clebsch-Gordan coefficients explicitly, we get
\begin{eqnarray}
\hat{T}^{(0)}_{0}&=& -\frac{1}{\sqrt{3}}\left( \dZero \dZeroDagger - \dPlus \dMinusDagger
    - \dMinus \dPlusDagger \right)\\
\hat{T}^{(1)}_{0}&=&\frac{1}{\sqrt{2}}\left(\dPlus \dMinusDagger - \dMinus \dPlusDagger \right)\nonumber\\
\hat{T}^{(1)}_{+1}&=&\frac{1}{\sqrt{2}}\left(-\dZero \dPlusDagger + \dPlus \dZeroDagger \right)\nonumber\\
\hat{T}^{(1)}_{-1}&=&\frac{1}{\sqrt{2}}\left(\dZero \dMinusDagger - \dMinus \dZeroDagger \right)\nonumber\\
\hat{T}^{(2)}_{0}&=&\frac{1}{\sqrt{6}}\left(\dPlus \dMinusDagger  +2 \dZero\dZeroDagger + \dMinus \dPlusDagger \right)\nonumber\\
\hat{T}^{(2)}_{+1}&=&\frac{1}{\sqrt{2}}\left(\dZero \dPlusDagger + \dPlus \dZeroDagger \right)\nonumber\\
\hat{T}^{(2)}_{-1}&=&\frac{1}{\sqrt{2}}\left(\dZero \dMinusDagger + \dMinus \dZeroDagger \right)\nonumber\\
\hat{T}^{(2)}_{+2}&=&\dPlus \dPlusDagger\nonumber\\
\hat{T}^{(2)}_{-2}&=&\dMinus \dMinusDagger\nonumber
\end{eqnarray}
Note that several standard references (including references \cite{Baym1969, Sakurai1994}) contain an error in the prefactor of the $j=0$ term and in the sign of the $j=1$ terms.  However, the fundamental definitions of $\hat{T}^{(j)}_m$ and its inverse above are valid.

Using recursion relations for the Clebsch-Gordan coefficients we can recast the tensor operators in terms of more intuitive $\hat{f}$ operators \cite{Polzik2004, Varshalovich1988}
\begin{eqnarray}
\hat{T}^{(0)}_{0}&=& -\alphaZero \fIdent/\sqrt{3} \label{Equation::T_f}\\
\hat{T}^{(1)}_{0}&=& +\alphaOne \fZ/\sqrt{2}\nonumber\\
\hat{T}^{(1)}_{+1}&=& +\alphaOne \fPlus/\sqrt{2}\nonumber\\
\hat{T}^{(1)}_{-1}&=& +\alphaOne \fMinus/\sqrt{2}\nonumber\\
\hat{T}^{(2)}_{0}&=& -\alphaTwo\left(3\fZSquared - f(f+1)\fIdent\right)/\sqrt{6}\nonumber\\
\hat{T}^{(2)}_{+1}&=& -\alphaTwo\sqrt{2}\fPlus\left(\fZ+\fIdent/2 \right)\nonumber\\
\hat{T}^{(2)}_{-1}&=& -\alphaTwo\sqrt{2}\fMinus\left(\fZ-\fIdent/2\right)\nonumber\\
\hat{T}^{(2)}_{+2}&=& -\alphaTwo \fPlusSquared\nonumber\\
\hat{T}^{(2)}_{-2}&=& -\alphaTwo \fMinusSquared\nonumber
\end{eqnarray}
Here we have defined
\begin{eqnarray}  \label{Equation::AlphaZero}
  \alphaZero & = & \alphaConst{f}{f'}\left( (2f-1)\kron{f-1}{f'} \right.+(2f+1)\kron{f}{f'}\nonumber\\
  & &\left. +(2f+3)\kron{f+1}{f'}\right)\\
  \label{Equation::AlphaOne}
  \alphaOne & = & \alphaConst{f}{f'}\left(-\frac{2f-1}{f}\kron{f-1}{f'} \right. -\frac{2f+1}{f(f+1)}\kron{f}{f'}\nonumber\\
  && \left. +\frac{2f+3}{f+1}\kron{f+1}{f'}\right)\\
  \label{Equation::AlphaTwo}
  \alphaTwo & = & \alphaConst{f}{f'}\left(\frac{1}{f}\kron{f-1}{f'} \right. -\frac{2f+1}{f(f+1)}\kron{f}{f'}\nonumber\\
  && \left. +\frac{1}{f+1}\kron{f+1}{f'}\right)
\end{eqnarray}
These definitions have been chosen to make the each of the quantities
\begin{equation}
\sum_{f^\prime} \frac{\alpha^{(j)}_{f,f^\prime}}{\alpha_0 \Delta_{f,f^\prime}} > 0
\end{equation}
for $\Delta_{f,f^\prime}\gg0$ for each term $j$.  We have defined the polarizability constants
\begin{eqnarray}
   \alphaConst{f}{f^\prime} &=& \alpha_0 \frac{(2 j^\prime+1)^2}{(2j+1)^2} \left|   \left\{
    \begin{array}{ccc}
       1 & j & j^\prime \\ i & f^\prime & f
    \end{array}  \right\} \right|^2
\end{eqnarray}
and
\begin{eqnarray} \label{Equation::AlphaConstant}
    \alpha_0  &=& \frac{3 \epsilon_0 \hbar \Gamma \lambda_0^3}{8 \pi^2} \\
    &=&   \left |  \langle j || \dVec || j' \rangle \right |^2 \frac{(2 j+1)}{(2j^\prime+1)} \nonumber
\end{eqnarray}
which involves the atomic the spontaneous emission rate, $\Gamma$, and transition wavelength, $\lambda_0$.

Now, to complete the derivation, insert Eqs. (\ref{Equation::T_f}) into the polarizability components of Eqs. (\ref{Equation::alphaExpand}), then insert this and the definition of the electric field, Eq. (\ref{Equation::EField}), into the Hamiltonian, Eqs. (\ref{Equation::HDecomp}-\ref{Equation::HComponent}).  Expanding, using the properties of the spherical dot product, and the Stokes component definitions (Eqs. \ref{Equation::Stokes}), and summing over the $f'$, we get the final expressions used in the text (Eqs. \ref{Equation::ScalarHamiltonian}, \ref{Equation::VectorHamiltonian}, \ref{Equation::TensorHamiltonian}).

\section{Arbitrary Vector Operator Rotations}\label{Section::RotateApp}

Here we are interested in evaluating the general operation of rotating a vector about an arbitrary direction by an arbitrary amount in order to determine the semiclassical evolution of the probe light as used in Eqs. (\ref{Equation::RotateS}).

Consider the rotation of the vector spin operator
\begin{equation}
\vec{S} = \left[\hat{S}_x, \hat{S}_y, \hat{S}_z\right]
\end{equation}
in Cartesian coordinates about an arbitrary direction $\vec{n}=[\gamma_x, \gamma_y, \gamma_z]/\gamma$ by the angle $\gamma=\sqrt{\gamma_x^2+\gamma_y^2+\gamma_z^2}$.  This rotation can be represented in the Heisenberg picture as
\begin{equation}
\hat{S}'_i = \hat{U} \hat{S}_i \hat{U}^{\dagger}
\end{equation}
where
\begin{equation}
\hat{U}=\exp[-i \gamma \vec{S}\cdot\vec{n}]=\exp[-i(\gamma_x \hat{S}_x +\gamma_z \hat{S}_z + \gamma_z \hat{S}_z)]
\end{equation}
%Use Eq. 3.16 of [http://www.phys.vt.edu/~mizutani/quantum/rotations.pdf]
The $\hat{S}'_i$ can be derived explicitly using the following equation for the arbitrary rotation of any vector
\begin{eqnarray}
\hat{S}'_i &=&(\vec{S}  \cdot \vec{i}) \cos\gamma + (\vec{n} \cdot \vec{i}) (\vec{n} \cdot \vec{S}) (1-\cos\gamma) \nonumber\\
 &&+ \left((\vec{n} \times\vec{i}) \cdot \vec{S}\right) \sin\gamma
\end{eqnarray}
Expanding and rearranging terms we get
\begin{eqnarray}
\hat{S}'_x &=& \hat{S}_x \left(\frac{\gamma_x^2}{\gamma^2}(1-\cos\gamma) + \cos\gamma\right)\nonumber\\
&&+ \hat{S}_y \left(\frac{\gamma_x\gamma_y}{\gamma^2}(1-\cos\gamma) + \frac{\gamma_z}{\gamma}\sin\gamma\right)\nonumber\\
&&+ \hat{S}_z \left(\frac{\gamma_x\gamma_z}{\gamma^2}(1-\cos\gamma) - \frac{\gamma_y}{\gamma}\sin\gamma\right)\\
\hat{S}'_y &=& \hat{S}_x \left(\frac{\gamma_y\gamma_x}{\gamma^2}(1-\cos\gamma) - \frac{\gamma_z}{\gamma}\sin\gamma\right)\nonumber\\
&& + \hat{S}_y \left(\frac{\gamma_y^2}{\gamma^2}(1-\cos\gamma) + \cos\gamma\right)\nonumber\\
&& + \hat{S}_z \left(\frac{\gamma_y\gamma_z}{\gamma^2}(1-\cos\gamma) + \frac{\gamma_x}{\gamma}\sin\gamma\right)\\
\hat{S}'_z &=& \hat{S}_x \left(\frac{\gamma_z\gamma_x}{\gamma^2}(1-\cos\gamma) + \frac{\gamma_y}{\gamma}\sin\gamma\right)\nonumber\\
&& + \hat{S}_y \left(\frac{\gamma_z\gamma_y}{\gamma^2}(1-\cos\gamma) - \frac{\gamma_x}{\gamma}\sin\gamma\right)\nonumber\\
&&  + \hat{S}_z \left(\frac{\gamma_z^2}{\gamma^2}(1-\cos\gamma) + \cos\gamma\right)
\end{eqnarray}
These equations can be specialized to Eqs. (\ref{Equation::RotateS}) which describes the experimental situation considered in this work.

\end{document}